\documentclass[reqno]{article}

\usepackage{graphicx}
\usepackage{amsmath,amsthm,amssymb}

\chardef\bslash=`\\ 

\theoremstyle{definition}

\theoremstyle{remark}

\newcommand{\eval}[2][\right]{\relax
  \ifx#1\right\relax \left.\fi#2#1\rvert}

\begin{document}
\title{\bf{On convergence towards a self-similar solution for a nonlinear wave
equation - a case study}}

\author{Piotr Bizo\'n\footnotemark[1]{}\; and
  Tadeusz Chmaj\footnotemark[2]{}\;\,\footnotemark[3]{}\;\\
  \footnotemark[1]{} \small{\textit{M. Smoluchowski Institute of Physics,
   Jagellonian University, Reymonta 4, 30-059 Krak\'ow, Poland}}\\
   \footnotemark[2]{} \small{\textit{H. Niewodniczanski Institute of Nuclear
   Physics,
Polish Academy of Sciences, Krak\'ow,
    Poland}}\\ \footnotemark[3]{} \small{\textit{Cracow University of Technology, Warszawska 24, 31-155 Krak\'ow,
    Poland}}\\}
%
\maketitle
\begin{abstract}
\noindent We consider the problem of asymptotic stability of a
self-similar attractor for a simple semilinear radial wave
equation which arises in the study of  the Yang-Mills equations in
5+1 dimensions. Our analysis consists of two steps. In the first
step we determine the spectrum of linearized perturbations about
the attractor using a method of continued fractions. In the second
step
 we demonstrate numerically that the resulting
eigensystem provides an accurate description of the dynamics of
convergence towards the attractor.
\end{abstract}
\section{Introduction}
Self-similar solutions of evolution equations often appear as
attractors in a sense that solutions of an initial value problem
starting from generic initial data evolve asymptotically into a
self-similar form. In such cases one would like to describe the
process of convergence to the self-similar solution and understand
the mechanism responsible for this phenomenon. These kind of
problems are relatively well-understood for diffusion equations
where the global dissipation of energy is the mechanism of
convergence to an attractor, however very little is known for
conservative wave equations where the local dissipation of energy
is due to dispersion. In this paper we report on analytical and
numerical studies of this problem for a semilinear radial wave
equation of the form
\begin{equation}\label{main}
u_{tt} - u_{rr} - \frac{2}{r} u_r + \frac{f(u)}{r^2}=0,
\end{equation}
 where $r$ is the radial variable, $u=u(t,r)$, and $f(u)=-3 u (1-u^2)$. This equation appears in the study of
the spherically symmetric Yang-Mills equations in $5+1$ dimensions
(see \cite{acta} for the derivation). We remark in passing that
our results hold for more general nonlinearities, in particular
for $f(u)= \sin(2u)$ which corresponds to the equivariant wave
maps from the $3+1$ dimensional Minkowski spacetime into the
three-sphere.

It was proved in \cite{cst} and later found explicitly in
\cite{acta} that equation (\ref{main}) has a self-similar solution
\begin{equation}\label{css0}
    u(t,r)=U_0(\rho)=\frac{1-\rho^2}{1+\frac{3}{5}\rho^2},
\end{equation}
where $\rho=r/(T-t)$ is a similarity variable and $T>0$ is a
constant (actually, $U_0$ is the ground state of a countable
family of self-similar solutions $U_n$ ($n=0,1,\dots$) but since
all $n>0$ solutions are unstable they do not appear as attractors
for generic initial data). Since
\begin{equation}\label{ssblow}
  \partial_r^2 U_0(\rho)\Bigr\rvert_{r=0} \sim \frac{1}{(T-t)^2},
\end{equation}
the solution $U_0(\rho)$ becomes singular at the center when $t
\rightarrow T$. By the finite speed of propagation, one can
    truncate this solution in space to get a smooth solution with compactly supported
    initial data which blows up in finite time.

In fact, the self-similar solution $U_0$ is not only an explicit
example of singularity formation, but more importantly it appears
as an attractor in the dynamics of generic initial data. We
conjectured in \cite{bt} that generic solutions of equation
(\ref{main}) starting with sufficiently large initial data do blow
up in a finite time in the sense that $u_{rr}(t,0)$ diverges at $t
\nearrow T$ for some $T>0$ and the asymptotic profile of blowup is
universally given by $U_0$, that is
\begin{equation}\label{conj}
 \lim_{t\nearrow T} u(t,(T-t) r) = U_0(r).
\end{equation}
Figure 1 shows the numerical evidence supporting this conjecture.
\begin{figure} [h]
\centering
\includegraphics[width=0.55\textwidth]{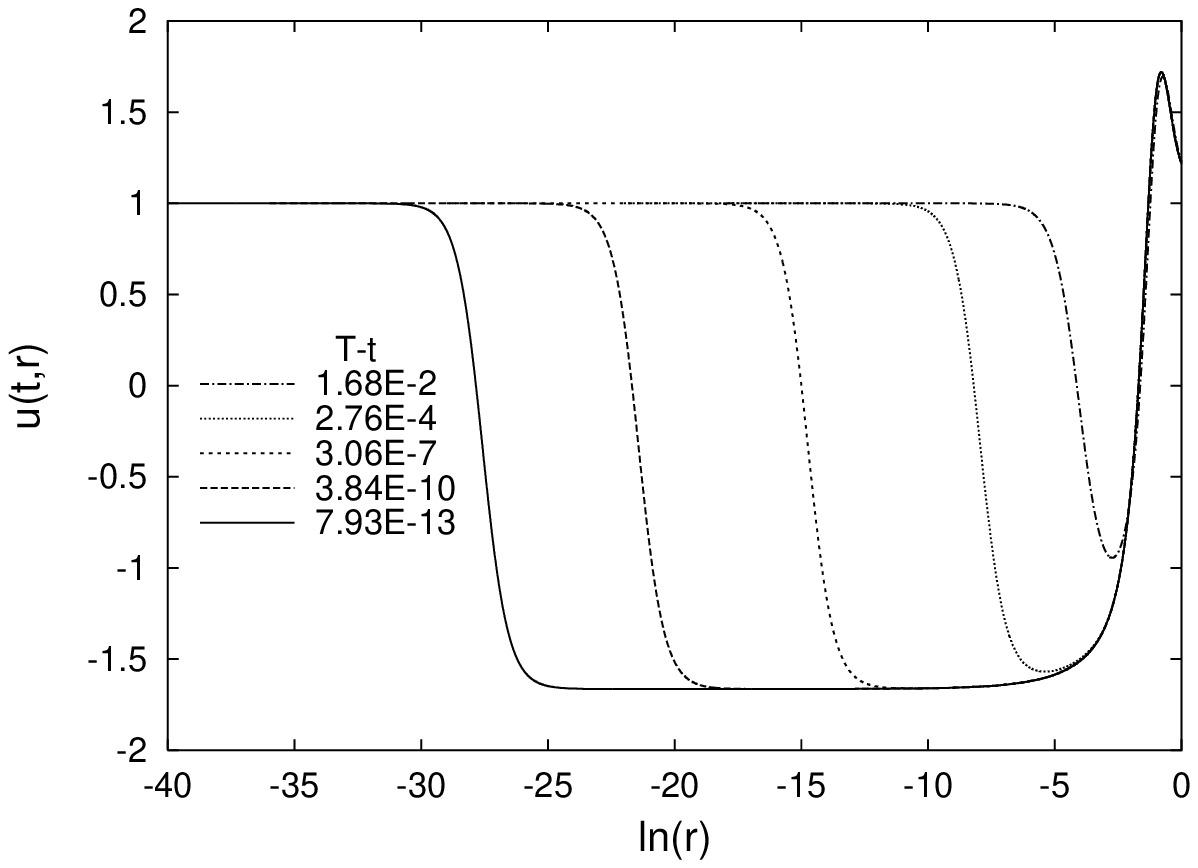}
\includegraphics[width=0.55\textwidth]{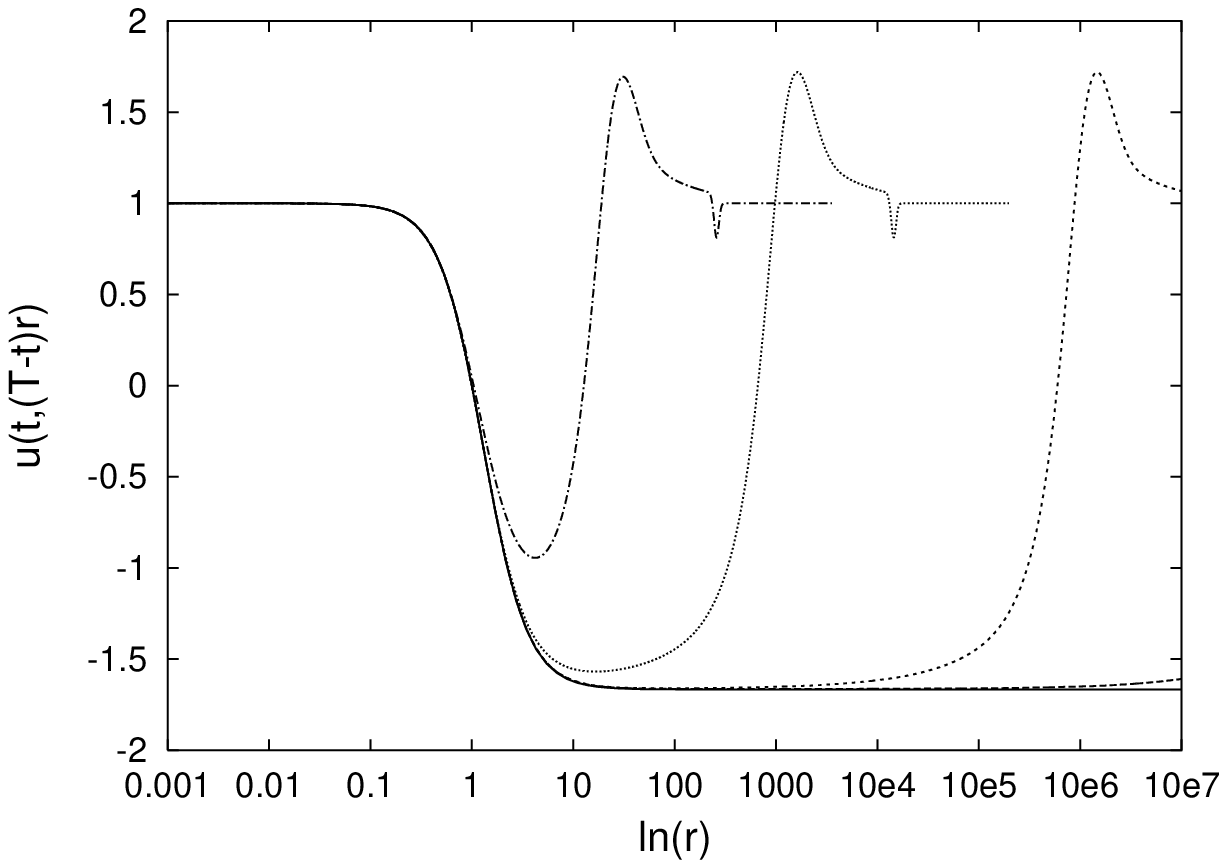}
\caption{\small{The upper plot shows the late time evolution of
some large initial data which blow up at time $T$. As the blowup
progresses, the inner solution gradually attains
 the form of the stable self-similar solution $U_0(r/(T-t))$. The outer solution appears frozen
on this timescale. In the lower plot the rescaled solutions $u(t,
(T-t)r)$ are shown to collapse  to the profile $U_0(r)$ (solid
line).} }\label{fig1}
\end{figure}

\noindent The goal of this paper is to describe in detail how the
limit (\ref{conj}) is attained. To this end, in section 2 we first
study the linear stability of the solution $U_0$. This leads to an
eigenvalue problem which is rather unusual from the standpoint of
spectral theory of linear operators. We solve this problem in
section 3 using the method of continued fractions. Then, in
section 4 we present the numerical evidence
 that the deviation of the dynamical solution from the
self-similar attractor is asymptotically well described by the
least damped eigenmode.
\section{Linear stability analysis}
The role of the self-similar solution $U_0$ in the evolution
depends crucially on its stability with respect to small
perturbations.
 In order to analyze this problem it is convenient to define the new time coordinate
$\tau=-\ln(T-t)$ and rewrite equation (\ref{main}) in terms of
$U(\tau,\rho)=u(t,r)$
\begin{equation} \label{rho-tau}
U_{\tau\tau} + U_{\tau} + 2 \rho\: U_{\rho\tau}
-(1-\rho^2)(U_{\rho\rho} +\frac{2}{\rho} U_{\rho})
+\frac{f(U)}{\rho^2}  = 0.
\end{equation}
In these variables the problem of finite time blowup in converted
into the problem of asymptotic convergence for $\tau \rightarrow
\infty$ towards the stationary solution $U_0(\rho)$.
 Following the standard procedure we seek
solutions of equation (\ref{rho-tau}) in the form
$U(\tau,\rho)=U_0(\rho)+ w(\tau,\rho)$. Neglecting the $O(w^2)$
terms we obtain a linear evolution equation for the perturbation
$w(\tau,\rho)$
\begin{equation}\label{pert}
w_{\tau\tau} + w_{\tau} + 2 \rho\: w_{\rho\tau}
-(1-\rho^2)(w_{\rho\rho} +\frac{2}{\rho} w_{\rho})
+\frac{f'(U_0)}{\rho^2} w  = 0.
\end{equation}
Substituting $w(\tau,\rho)=e^{\lambda \tau} v(\rho)/\rho$ into
(\ref{pert}) we get the eigenvalue equation
\begin{equation}\label{spectrum}
-(1-\rho^2) v''+2 \lambda \rho v' +\lambda(\lambda-1) v +
\frac{V(\rho)}{\rho^2} v=0,
\end{equation}
where
\begin{equation}\label{poten}
    V(\rho) = f'(U_0)= \frac{6 (25-90\rho^2+33\rho^4)}{(5+3 \rho^2)^2}.
\end{equation}
We consider equation (\ref{spectrum}) on the interval $0\leq
\rho\leq 1$, which corresponds to the interior of the past light
cone of the blowup point $(t=T,r=0)$. Since a
 solution of  the initial value problem for equation
(\ref{main}) starting from smooth initial data remains smooth for
all times $t<T$, we demand the solution $v(\rho)$ to be analytic
at the both endpoints $\rho=0$ (the center) and $\rho=1$ (the past
light cone). Such a globally analytic solution of the singular
boundary value problem can exist only for discrete values of the
parameter $\lambda$, hereafter called eigenvalues. In order to
find the eigenvalues we apply the method of Frobenius.

The indicial exponents at the regular singular point $\rho=0$ are
$3$ and $-1$, hence the solution which is analytic at $\rho=0$ has
the power series representation
\begin{equation}\label{v0}
v_0(\rho) = \sum_{n=0}^{\infty} a_n(\lambda) \rho^{2n+3}, \qquad
a_0 \neq 0.
\end{equation}
Since the nearest singularity in the complex $\rho$-plane is at
$\rho=1$, the series (\ref{v0}) is absolutely convergent for
$0\leq \rho < 1$. At the second regular singular point, $\rho=1$,
the indicial exponents are $0$ and $1-\lambda$ so, as long as
$\lambda$ is not an integer, the two linearly independent
solutions have the power series representations
\begin{equation}\label{v12}
v_1(\rho) = \sum_{n=0}^{\infty} b_n^{(1)}(\lambda)
(1-\rho)^n,\qquad v_2(\rho) = \sum_{n=0}^{\infty}
b_n^{(2)}(\lambda) (1-\rho)^{n+1-\lambda}.
\end{equation}
These series are absolutely convergent for $0< \rho \leq 1$.
 If $\lambda$ is not an integer,
only the solution $v_1(\rho)$ is  analytic at $\rho=1$. From the
theory of linear ordinary differential equations we know that the
three solutions $v_0(\rho)$, $v_1(\rho)$, and $v_2(\rho)$ are
connected  on the interval $0<\rho<1$ by the linear
relation\footnote{If $1-\lambda$ is a positive integer $N$, then
the solution which is analytic at $\rho=1$ behaves as $v_1\sim
(1-\rho)^{n+N}$ while the second solution $v_2$  involves the
logarithmic term $C_N v_1 \ln(1-\rho)$. By a straightforward but
tedious calculation one can check that the coefficient $C_N$ is
nonzero for all $N$.}
\begin{equation}\label{connect}
    v_0(\rho) = A(\lambda) v_1(\rho) + B(\lambda) v_2(\rho).
\end{equation}
The requirement that the solution which is analytic at $\rho=0$ is
also analytic at $\rho=1$ serves as the quantization condition for
the eigenvalues $B(\lambda)=0$. Unfortunately, the explicit
expressions for the connection coefficients $A(\lambda)$ and
$B(\lambda)$ are not known for equations with more than three
regular singular points. There are, however, other indirect
methods of solving the equation $B(\lambda)=0$. One of them is a
shooting method which goes as follows. One approximates the
solutions $v_0(\rho)$ and $v_1(\rho)$ by the power series
(\ref{v0}) and (\ref{v12}), respectively, truncated at some
sufficiently large $n$, and then computes the Wronskian of these
solutions at  midpoint, $\rho=1/2$, say. The zeros of the
Wronskian correspond to the eigenvalues (see figure 2). Although
this method gives the eigenvalues with reasonable accuracy, it is
computationally very costly, especially for large negative values
of $\lambda$, because the power series (\ref{v0}) and (\ref{v12})
converge very slowly. We point out that shooting towards $\rho=1$
fails completely for large negative $\lambda$ because the solution
$v_2(\rho)$ is subdominant at $\rho=1$, that is, it is negligible
with respect to the analytic solution $v_1(\rho)$.
\begin{figure} [hb]
\centering
\includegraphics[width=0.7\textwidth, angle=-90]{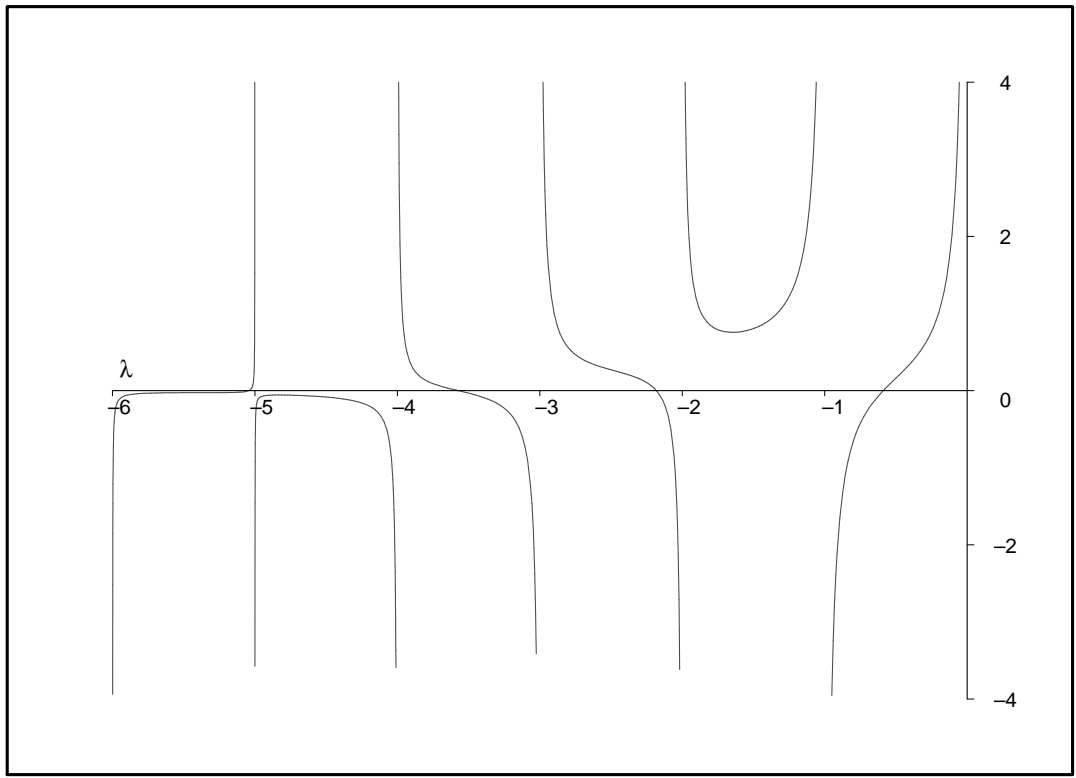}
\caption{\small{The Wronskian $W[v_0,v_1](\rho=1/2,\lambda)$.}
}\label{fig2}
\end{figure}
\section{The continued fractions method}
In this section we shall solve
 the eigenvalue problem (\ref{spectrum}) using a method
continued fractions.
 The key
idea is to determine the analyticity properties of the power
series solution $v_0(\rho)$ from the asymptotic behavior of the
expansion coefficients $a_n$. Substituting the power series
(\ref{v0}) into equation (\ref{spectrum}) we get the four-term
recurrence relation (with the initial conditions $a_0=1$
(normalization) and $a_n=0$ for $n<0$)
\begin{equation}\label{4term}
    p_3(n) a_{n+3} + p_2(n) a_{n+2} + p_1(n) a_{n+1} + p_0(n) a_n
    = 0,
\end{equation}
where
\begin{eqnarray*}
p_3(n) & =& - 100 n^2 - 850 n - 1650,\\
p_2(n) &=& -20 n^2 + (100\lambda-130) n +\! 25\lambda^2 + 325\lambda -750,\\
p_1(n) &=& 84 n^2 + (120\lambda+378) n + 30\lambda^2+270\lambda+618,\\
p_0(n) & =& 36 n^2 + (36\lambda+90)n + 9\lambda^2 + 45\lambda +54.
\end{eqnarray*}
For $n=-2$ we have
\begin{equation}\label{4term-2}
 -350 a_1 + (25\lambda^2+125\lambda-570) a_0=0,
\end{equation}
and for $n=-1$
 \begin{equation}
 -900 a_2 +
(25\lambda^2+225\lambda-640) a_1 + (30\lambda^2+150\lambda+324)
a_0 = 0.
\end{equation}
The series (\ref{v0}) is absolutely convergent for $0\leq \rho<1$
and in general is divergent for $\rho>1$.
 In
order to determine the analyticity properties of the solution
$v_0(\rho)$ at $\rho=1$ we need to find the large $n$ behavior of
the expansion coefficients $a_n$. The four-term recurrence
relation (\ref{4term}) can be viewed as the third order difference
equation so it has three linearly independent asymptotic solutions
for $n\rightarrow\infty$. Following standard methods (see, for
example, \cite{de}) we find
\begin{equation}\label{asym4}
a_n^{(1)} \sim n^{\lambda-2} \sum_{s=0}^{\infty}
\frac{\alpha_s^{(1)}}{n^s},\qquad a_n^{(2)} \sim
\left(-\frac{3}{5}\right)^n n \sum_{s=0}^{\infty}
\frac{\alpha_s^{(2)}}{n^s}, \qquad a_n^{(3)} \sim
\left(-\frac{3}{5}\right)^n n^{-4} \sum_{s=0}^{\infty}
\frac{\alpha_s^{(3)}}{n^s}.
\end{equation}
Thus, in general, the solution of the recurrence relation
(\ref{4term}) behaves asymptotically as
\begin{equation} \label{4termasym}
a_n \sim c_1(\lambda) a_n^{(1)}+c_2(\lambda)
a_n^{(2)}+c_3(\lambda) a_n^{(3)}.
 \end{equation}
  If the
coefficient $c_1(\lambda)$ is nonzero then
\begin{equation}
\frac{a_{n+1}}{a_n} \sim
\frac{a^{(1)}_{n+1}}{a^{(1)}_n}\rightarrow 1 \quad \text{as}\quad
n\rightarrow\infty,
\end{equation}
hence the power series (\ref{v0}) is divergent for $\rho>1$ (in
fact it has a branch point singularity at $\rho=1$). On the other
hand, if $c_1(\lambda)=0$ then the solution $v_0(\rho)$ can be
continued analytically through $\rho=1$.
 The
advantage of replacing the quantization condition $B(\lambda)=0$
in the connection formula (\ref{connect})  by the equivalent
condition
 $c_1(\lambda)=0$ follows from the fact that $c_1(\lambda)$ is the coefficient of
the dominant solution in (\ref{4termasym}), in contrast to
$B(\lambda)$ which is the coefficient of the subdominant solution
in (\ref{connect}).

Now, we shall find the zeros of the coefficient $c_1(\lambda)$
using the method of continued fractions. This method is based on
an intimate relationship between three-term recurrence relations
and continued fractions. It goes as follows. Suppose that we have
a three-term recurrence relation (a second order difference
equation)
\begin{equation}\label{gen3term}
b_{n+2} + A_n b_{n+1} + B_n b_n =0.
\end{equation}
Let $r_n=b_{n+1}/b_n$. Then, from (\ref{gen3term}) we have
$$
r_n = - \frac{B_n}{A_n+r_{n+1}},
$$
and applying this formula repeatedly we get the continued fraction
representation of $r_n$
\begin{equation}\label{pinch}
r_n = - \frac{B_n}{A_n -} \;\frac{B_{n+1}}{A_{n+1} -}\;
\frac{B_{n+2}}{A_{n+2} -} ...
\end{equation}
A theorem due to Pincherle \cite{de} says that the continued
fraction on the right hand side of equation (\ref{pinch})
converges if and only if the recurrence relation (\ref{gen3term})
has a minimal solution $b^{min}_n$, i.e. the solution such that
$\lim_{n \rightarrow \infty} b^{min}_n/b_n=0$ for any other
solution $b_n$. Moreover, in the case of convergence, equation
(\ref{pinch}) holds with $r_n=b^{min}_{n+1}/b^{min}_n$ for each
$n$.

We cannot yet apply this theorem because our recurrence relation
(\ref{4term}) has four terms. However, let us observe that
\begin{equation}
a_n=a_n^{exact}=\left(-\frac{3}{5}\right)^n \left[n+1+\frac{5}{16}
(\lambda-1)\right]
\end{equation}
is the exact solution of our four-term recurrence relation
(\ref{4term}) (although it does not satisfy the initial
conditions).
 Using
this solution we can reduce the order by the substitution
\begin{equation}\label{bn}
b_n=a_{n+1}+ \frac{3}{5} \frac{n+2+\frac{5}{16} (\lambda-1)}{
n+1+\frac{5}{16} (\lambda-1)}\;a_n
\end{equation}
 to get the three-term recurrence relation
 \begin{equation}\label{3term}
    q_2(n) b_{n+2} + q_1(n) b_{n+1} + q_0(n) b_n
    = 0,
\end{equation}
where (using the abbreviation $\gamma=5 (\lambda-1)/16$)
\begin{eqnarray*}
q_0(n) &= & p_1(n) - \frac{3}{5} \frac{n+3+\gamma}{n+2+\gamma}\,
p_2(n)+
\left(\frac{3}{5}\right)^2 \frac{n+4+\gamma}{n+2+\gamma}\,p_3(n), \\
q_1(n) &=&p_2(n)- \frac{3}{5} \frac{n+4+\gamma}{n+3+\gamma}\, p_3(n),\\
 q_2(n) &=& p_3(n).
\end{eqnarray*}
The two linearly independent asymptotic solutions of the
recurrence relation (\ref{3term}) are
\begin{equation}\label{bnasym}
b_n^{(1)} \sim n^{\lambda-2} \sum_{s=0}^{\infty}
\frac{\beta_s^{(1)}}{n^s}, \qquad a_n^{(2)} \sim
\left(-\frac{3}{5}\right)^n n^{-5} \sum_{s=0}^{\infty}
\frac{\beta_s^{(2)}}{n^s},
\end{equation}
so in general
\begin{equation}
 b_n \sim C_1(\lambda) b_n^{(1)}+C_2(\lambda)
b_n^{(2)}.
\end{equation}
 Now, our quantization condition $c_1(\lambda)=0$ is equivalent to $C_1(\lambda)=0$ which is
 nothing else but  the condition for the existence of a minimal
 solution for equation (\ref{3term}).
Thus, we can use  Pincherle's theorem to find the eigenvalues.

In our case $A_n=q_1(n)/q_2(n)$ and $B_n=q_0(n)/q_2(n)$. From
(\ref{4term-2}) and (\ref{bn}) we get \begin{equation} b_0=
\left(\frac{25\lambda^2+125\lambda-570}{350}+ \frac{96+
15(\lambda-1)}{80+ 25(\lambda-1)}\right) a_0 ,\qquad b_{-1}=a_0.
\end{equation}
 Using  Pincherle's theorem and setting $n=-1$ in
 (\ref{pinch}) we obtain the eigenvalue equation
\begin{equation}\label{eigen}
\frac{b_0}{b_{-1}} = \frac{25\lambda^2+125\lambda-570}{350}+
\frac{96+ 15(\lambda-1)}{80+ 25 (\lambda-1)} = -
\frac{B_{-1}(\lambda)}{A_{-1}(\lambda) -}\;
\frac{B_{0}(\lambda)}{A_{0}(\lambda) -}
\;\frac{B_{1}(\lambda)}{A_{1}(\lambda) -} ...
\end{equation}
The continued fraction in (\ref{eigen}), which by Pincherle's
theorem is convergent for any $\lambda$, can be approximated with
essentially arbitrary accuracy  by downward recursion starting
from a sufficiently large $n=N$ and some (arbitrary) initial value
$r_N$. The roots of the transcendental equation (\ref{eigen}) are
then found numerically (see table 1).
\begin{table}[h]
\centering
$$
\begin{tabular}{|c|cccccc|} \hline
$n$ & $0$ & $1$ & $2$ & $3$ & $4$ & $5$\\
& & & & & & \\
 $\lambda_n$ & 1 &  -0.588904  & -2.181597  & -
3.570756 & - 5.043294 & -6.486835\\
 \hline \hline

$n$ & $6$ & $7$ & $8$ & $9$ & $10$ & $11$\\
& & & & & &  \\
$\lambda_n$ & -7.912777 & -9.298265 & -9.907103& -10.792456 & -12.153033 & -13.164487\\
 \hline
\end{tabular}
$$
\caption{The first twelve  eigenvalues.}
\end{table}

\noindent We recall from \cite{acta} that the  eigenvalue
$\lambda_0=1$ is due to the freedom of changing the blowup time
$T$. Although this
 eigenvalue is positive, it should not be interpreted as the physical
instability of the solution $U_0$ -- it is an artifact of
introducing the similarity variables and does not show up in the
dynamics for $u(t,r)$. Note that, strangely enough, all the
eigenvalues are real.
\section{Convergence to the attractor}
According to the linear stability analysis presented above  the
convergence of the solution $u(t,r)$ towards the self-similar
attractor $U_0$ should be described by the formula
\begin{equation}\label{numer}
u(t,r) = U_0(\rho)+ \sum_{k=1}c_k e^{\lambda_k \tau}
v_k(\rho)/\rho \sim U_0(\rho)+c_1 e^{\lambda_1 \tau}
v_1(\rho)/\rho \quad \text{as} \quad \tau \rightarrow \infty,
\end{equation}
where $v_k(\rho)/\rho$ are the eigenmodes corresponding to the
eigenvalues $\lambda_k$ and $c_k$ are the expansion coefficients.
In order to verify the formula (\ref{numer})  we solved equation
(\ref{main}) numerically for large initial data leading to blowup,
expressed the solution in the similarity variables, and computed
the deviation from $U_0$ for $t \nearrow T$. Figure 4 shows that
for small $T-t$ the deviation from $U_0$ is very well described by
the least damped eigenmode $v_1$, in  agreement with the formula
(\ref{numer}). For larger $T-t$ the contribution of higher modes
has to be included (see figure 5).
\begin{figure}[h!]
\centering
\includegraphics[width=0.76\textwidth]{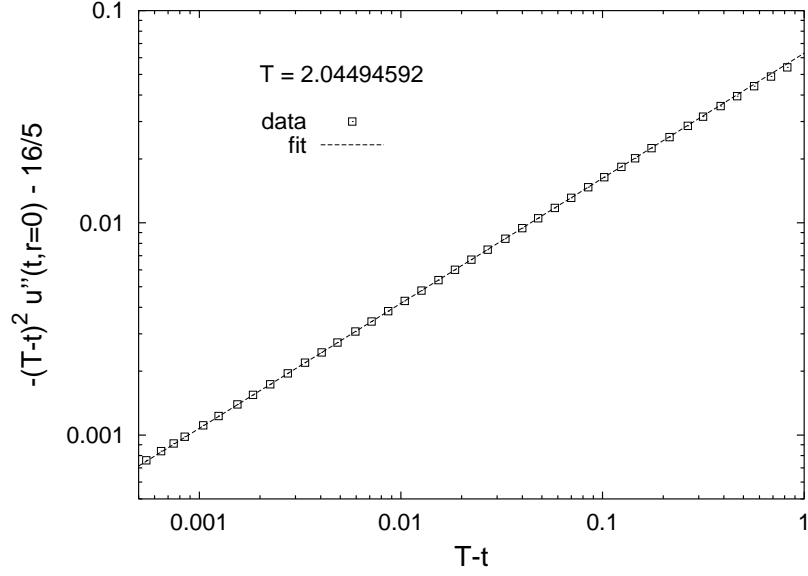}
\vskip 0.15cm
 \caption{\small{In this figure we determine the coefficient $c_1$
 in the formula (\ref{numer}) as follows. Taking the second
 derivative of (\ref{numer}) at $r=0$ we obtain
 $(T-t)^2 u_{rr}(t,r=0)+16/5=2 c_1 (T-t)^{-\lambda_1}$. For small $T-t$ we plot
 the left hand side of this equation in the log-log scale and fit
 the parameters $c_1$ and $\lambda_1$. We get $c_1=-0.031436$ and
 $\lambda_1=-0.58803$ (in good agreement with table 1).}}
\end{figure}
\begin{figure}[h!]
\centering
\includegraphics[width=0.76\textwidth]{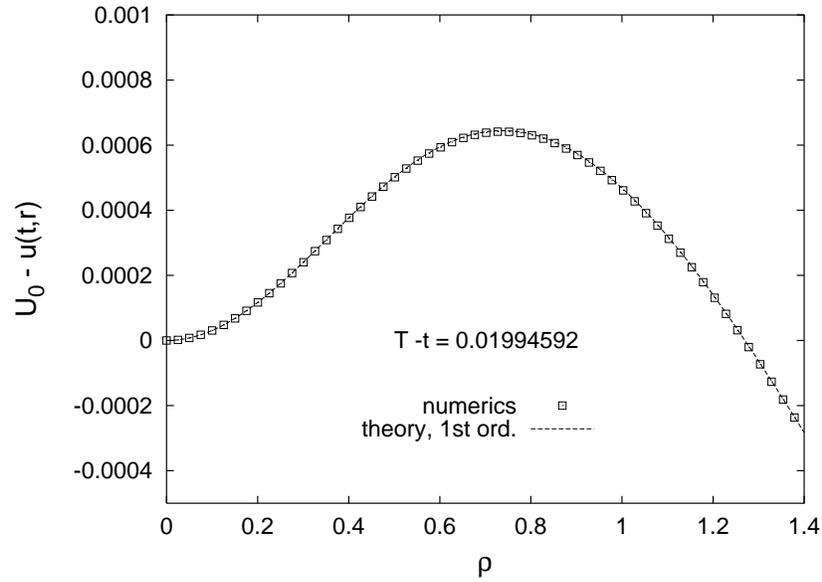}
\vskip 0.15cm
 \caption{\small{We plot the deviation of the dynamical solution
$u(t,r)$ from the self-similar solution $U_0$ at some moment of
time close to the blowup time. The solid line shows the least
damped eigenmode $c_1 (T-t)^{-\lambda_1} v_1(\rho)/\rho$ with the
coefficient $c_1$ obtained in figure 3.}}
\end{figure}
\begin{figure}[h!]
\centering
\includegraphics[width=0.76\textwidth]{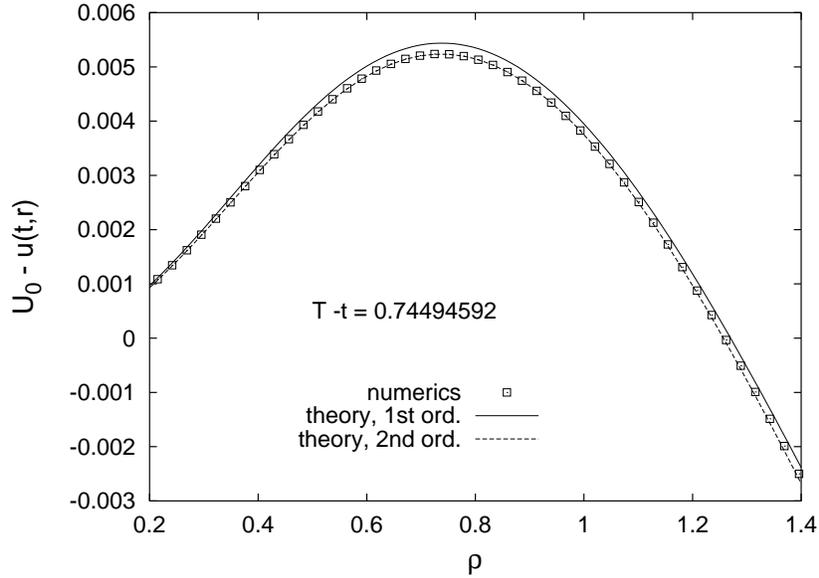}
\vskip 0.15cm
 \caption{\small{The same plot as in figure 4 but at an earlier
 moment of time. The solid line showing the least damped eigenmode
 (with the same coefficient $c_1$ as before) does not fit well the
 numerical data because the other modes have not yet decayed.  Including  the
 suitably normalized second eigenfunction in the expansion (\ref{numer}) we get a much better fit
 (dashed line).}}
\end{figure}

\newpage
\section*{Acknowledgments}
 This research  was supported in part
by the KBN grant 2 P03B 006 23. PB acknowledges the friendly
hospitality of the Albert Einstein Institute during part of this
work.

\end{document}